\documentclass[longauth,bibyear]{aa}
\usepackage{txfonts}
\usepackage{graphicx}   
\usepackage{amsmath}    
\usepackage[utf8]{inputenc} 
\usepackage{array} 
\usepackage{multicol,rotating} 
\usepackage{longtable} 
\usepackage[section]{placeins}
\providecommand{\tabularnewline}{\\}

\begin{document}

\newcommand{\abs}[1]{\ensuremath{\lvert#1\rvert}}

\title{Surface evolution of the Anhur region on comet 67P from high-resolution OSIRIS images}

\author{S. Fornasier\inst{1} 
\and C. Feller\inst{1} 
\and P.H. Hasselmann\inst{1} 
\and M.A. Barucci\inst{1} 
\and J. Sunshine\inst{2}
\and J.-B. Vincent\inst{3}
\and X. Shi\inst{4} 
\and H. Sierks\inst{4} 
\and G. Naletto\inst{5, 6, 7} 
\and P. L. Lamy\inst{8} 
\and R. Rodrigo\inst{9,10} 
\and D. Koschny\inst{11} 
\and B. Davidsson\inst{12} 
\and J.-L. Bertaux\inst{8} 
\and I. Bertini\inst{5} 
\and D. Bodewits\inst{13} 
\and G. Cremonese\inst{14} 
\and V.  Da Deppo\inst{7} 
\and S.  Debei\inst{15} 
\and M. De Cecco\inst{16} 
\and J. Deller\inst{4} 
\and S. Ferrari\inst{17} 
\and M. Fulle\inst{18} 
\and P. J. Gutierrez\inst{19} 
\and C. G\"uttler\inst{4} 
\and W.-H. Ip\inst{20,21} 
\and L. Jorda\inst{22}
\and H.U. Keller\inst{23, 24} 
\and M. L. Lara\inst{19} 
\and M. Lazzarin\inst{5} 
\and J.J. Lopez Moreno\inst{19} 
\and A. Lucchetti\inst{14} 
\and F. Marzari\inst{5} 
\and S. Mottola\inst{3} 
\and M. Pajola\inst{14} 
\and I. Toth\inst{25} 
\and C. Tubiana\inst{4}
}
%
\institute{LESIA, Observatoire de Paris, Universit\'e PSL, CNRS, Univ. Paris Diderot, Sorbonne Paris Cit\'{e}, Sorbonne Universit\'e, 5 Place J. Janssen, 92195 Meudon Pricipal Cedex, France \email{sonia.fornasier@obspm.fr}
\and Department of Astronomy, University of Maryland, College Park, MD 20742-2421, USA
\and Deutsches Zentrum f\"ur Luft und Raumfahrt (DLR), Institut f\"ur Planetenforschung, Asteroiden und Kometen, Rutherfordstrasse 2, 12489 Berlin, Germany
\and Max-Planck-Institut f\"ur Sonnensystemforschung, Justus-von-Liebig-Weg, 3, 37077, G\"ottingen, Germany
\and University of Padova, Department of Physics and Astronomy {\it Galileo Galilei}, Via Marzolo 8, 35131 Padova, Italy 
\and Center of Studies and Activities for Space (CISAS) {\it G. Colombo}, University of Padova, Via Venezia 15, 35131 Padova, Italy
\and CNR-IFN UOS Padova LUXOR, Via Trasea, 7, 35131 Padova, Italy
\and Laboratoire Atmosph\`eres, Milieux et Observations Spatiales, CNRS \& Universit\'e de Versailles Saint-Quentin-en-Yvelines, 11 boulevard d'Alembert, 78280 Guyancourt, France
\and Centro de Astrobiologia, CSIC-INTA, 28850 Torrejon de Ardoz, Madrid, Spain
\and International Space Science Institute, Hallerstrasse 6, 3012 Bern, Switzerland
\and Scientific Support Office, European Space Research and Technology Centre/ESA, Keplerlaan 1, Postbus 299, 2201 AZ Noordwijk ZH, The Netherlands
\and Jet Propulsion Laboratory, M/S 183-401, 4800 Oak Grove Drive, Pasadena, CA 91109, USA
\and Auburn University, Physics Department, 206 Allison Laboratory, Auburn, AL 36849, USA
\and INAF, Osservatorio Astronomico di Padova, Vicolo dell'Osservatorio 5, 35122 Padova, Italy
\and University of Padova, Department of Mechanical Engineering, via Venezia 1, 35131 Padova, Italy
\and University of Trento, Faculty of Engineering, Via Mesiano 77, 38100 Trento, Italy
\and Dipartimento di Geoscienze, University of Padova, via G. Gradenigo 6, 35131 Padova, Italy
\and INAF Astronomical Observatory of Trieste, Via Tiepolo 11, 34014 Trieste, Italy
\and Instituto  de Astrof\'isica de Andalucia (CSIC), c/ Glorieta de la Astronomia s/n, 18008 Granada, Spain
\and National Central University, Graduate Institute of Astronomy, 300 Chung-Da Rd, Chung-Li 32054, Taiwan
\and Space Science Institute, Macau University of Science and Technology, Avenida Wai Long, Taipa, Macau 
\and Laboratoire d'Astrophysique de Marseille, UMR 7326 CNRS, Universit\'e Aix-Marseille, 38 r
ue Fr\'ed\'eric Joliot-Curie, 13388 Marseille Cedex 13, France\
\and Institut f\"ur Geophysik und extraterrestrische Physik (IGEP), Technische Universitat Braunschweig, Mendelssohnstr. 3, 38106 Braunschweig, Germany
\and Operations Department, European Space Astronomy Centre/ESA, P.O.Box 78, 28691 Villanueva de la Canada, Madrid, Spain
\and MTA CSFK Konkoly Observatory, Budapest, Hungary 
}

\date{Accepted XXX. Received December 2018; in original form ZZZ}

\newpage

 \abstract{The southern hemisphere of comet 67P/Churyumov-Gerasimenko (67P) became observable by the Rosetta mission in March 2015, a few months before cometary southern vernal equinox. The Anhur region in the southern part of the comet's larger lobe was found to be highly eroded, enriched  in volatiles, and highly active.}{We analyze high-resolution images of the Anhur region pre- and post-perihelion acquired by the OSIRIS imaging system on board the Rosetta mission. The Narrow Angle Camera is particularly useful for studying the evolution in Anhur in terms of morphological changes and color variations.}{Radiance factor images processed by the OSIRIS pipeline were coregistered, reprojected onto the 3D shape model of the comet, and corrected for the illumination conditions.}{We find a number of morphological changes in the Anhur region that are related to formation of new scarps; removal of dust coatings; localized resurfacing in some areas, including boulders displacements; and vanishing structures, which implies localized mass loss that we estimate to be higher than 50 million kg. The strongest  changes took place in and nearby the Anhur canyon-like structure, where significant dust cover was removed, an entire structure vanished, and many boulders were rearranged. All such changes are potentially associated with one of the most intense outbursts registered by Rosetta during its observations, which occurred one day before perihelion passage. Moreover, in the niche at the foot of a new observed scarp, we also see evidence of water ice exposure that persisted for at least six months. The abundance of water ice, evaluated from a linear mixing model, is relatively high ($>$ 20\%). Our results confirm that the Anhur region is volatile-rich and probably is the area on 67P with the most pristine exposures near perihelion.}{}

\titlerunning{Anhur surface evolution}
   \maketitle

\section{Introduction}

\begin{table*}
         \begin{center} 
         \caption{Observing conditions for the NAC images ($\alpha$ is the  phase angle, r$_h$ is the heliocentric distance, and $\Delta$ is the distance between comet and spacecraft). The time refers to the start time of the first image of an observing  sequence, in case of multiple filter observations. Filters:  F22 (649.2 nm), F23 (535.7 nm), F24 (480.7 nm), F16 (360.0 nm), F27 (701.2 nm), F28 (743.7 nm), F41 (882.1 nm), F51 (805.3 nm), F61 (931.9 nm), F71 (989.3 nm), and F15 (269.3 nm). }
         \label{tab1}
        \begin{tabular}{|l|c|l|l|l|l|l|} \hline
time          & Filter & $\alpha$ & r$_{\odot}$ & $\Delta$ & res & figure  \\
              &        &   ($^{\circ}$)  & (au)        &   (km)   & (m/px) &       \\ \hline
2015-03-25T13h23 & F22, F23, F24, F41 & 73.2 & 2.017 & 87.7 & 1.70 & 2, 6, 7, 8, 10 \\
2015-05-02T10h42 & all &  61.5 & 1.732 &  125.0 &   2.4 & 3 \\
2015-10-31T15h07 & all  & 62.0 & 1.565 & 290.0 & 5.5 & 9 \\
2016-01-02T17h03 & F22                    & 90.0 & 2.03 & 83.2 &  1.6  &  4  \\
2016-01-27T18h36 & F22, F24, F41            & 62.0 & 2.23 & 70.0  &  1.4  &  4, 6  \\ 
2016-02-10T08h14 & F22, F23, F24, F16, F27, &  66.0  & 2.329 & 50.0   &   1.0 & 1, 2, 3, 7, 8, 10   \\
                & F28, F41, F51, F61, F71  &      &       &        &     &  \\

2016-06-25T01h37 &  F22, F23, F24, F16, F41 & 86.9 & 3.272 & 17.9 & 0.35 & 5, 11, 12, 13 \\

2016-06-25T11h50 & F22, F23, F24, F16, F41 & 86.7 & 3.272 & 15.71 & 0.30 & 11, 12, 13 \\
2016-07-30T05h02 & F22, F24, F41 & 99.7 & 3.486 &  9.04 & 0.17 & 15\\
2016-07-30T05h05 & F22, F24, F41 & 99.7 & 3.486 &  9.04 & 0.17 & 15\\
2016-07-30T05h09 & F22, F24, F41 & 99.7 & 3.486 &  9.04 & 0.17 & 15 \\
2016-07-30T05h12 & F22, F24, F41 & 99.7 & 3.486 &  9.04 & 0.17 & 15 \\ 
 2016-09-08T20h08 & F22                     &  93.1    &  3.72    &  4.03     & 0.08     & 6\\ 
2016-09-08T20h17 & F22                     &  93.1    &  3.72    &  4.03     & 0.08     & 5\\ 
\hline
        \end{tabular}
\end{center}
 \end{table*}
The Rosetta mission made unprecedented observations of the surface, activity, and evolution of comet 67P/Churyumov-Gerasimenko (67P hereafter) during about two years of continuous observation (Barucci \& Fulchignoni, 2017). During the extended mission, that is, from January to September 2016, when Rosetta was closer to the nucleus, it acquired high-resolution images of several regions of the comet. During this observing phase, Rosetta  in particular collected images  with sub-meter resolution of the southern hemisphere. The northern hemisphere had been imaged previously at high resolution during the 2014 nucleus mapping phase to identify and characterize the Philae landing site (Sierks et al., 2015, Thomas et al., 2015). In contrast, the southern hemisphere became visible from Rosetta only in March 2015, that is, two months before the southern vernal equinox. A clear morphological dichotomy was observed between the northern and southern hemispheres, with the latter showing a globally higher degree of erosion and a lack of wide-scale smooth terrains (El-Maarry et al., 2015a, 2016; Giacomini et al., 2016; Lee et al. 2016).\\ 

A number of localized morphological changes were reported for several regions of 67P (see El-Maarry et al. 2015a and 2016  for a definition of the regions on 67P). The reports suggested compositional and/or physical heterogeneity (El-Maarry et al., 2017).  A region with extensive changes is Imhotep, for which exhumation of structures (boulders and roundish features)  by the removal of $\sim$ 4 m dust coating (Auger et al. 2016, El-Maarry et al., 2017) and the appearance of two roundish structures with a diameter of $\sim$ 240 m and 140 m and a height of 5$\pm$2 m in the large smooth central area (Groussin et al., 2015) were reported. The transport of unconsolidated materials removes the dust coating; this is also observed on Anubis and Hapi (El-Maarry et al., 2017), the latter also showing aeolian-like ripples (Thomas et al., 2015). The two roundish features observed in Imhotep had a high expansion rate, higher than 18 cm hour$^{-1}$, that was not produced by the simple sublimation process. The features also hosted bright and bluer material at their edges, indicating exposed ice. The low surface tensile strength of the material together with the concurrence of exothermic processes such as the crystallization of water ice and/or the  clathrate destabilization have been invoked to explain the high expansion rate of the new observed structures in Imhotep (Groussin et al., 2015). \\ 
Extensive changes were also observed in the Aswan site in the Seth region. Aswan was one of the five selected potential landing sites of Philae (Pajola et al., 2016a). A cliff collapse produced a mass loss of  $\sim 10^{6}$ kg and exposed the water-ice-enriched inner layers of the comet (Pajola et al., 2017). This was the first observed link between an outburst and a cliff collapse on 67P.
In the Khonsu region, surface changes were also observed, including the 140 m displacement of a 30 m wide boulder (El-Maarry et al., 2017), the appearance of a new 50 meter-sized boulder (Hasselmann et al., 2019), several ice-enriched patches, one of which survived for several months (Deshapriya et al., 2016), and the sublimation of some thick dust layers (Hasselmann et al., 2019). However, even though numerous localized changes were reported, they did not substantially change the cometary landscape, which was very probably shaped much earlier in its history (El-Maarry et al., 2017). 

\begin{figure*}
\centering
\includegraphics[width=0.9\textwidth,angle=0]{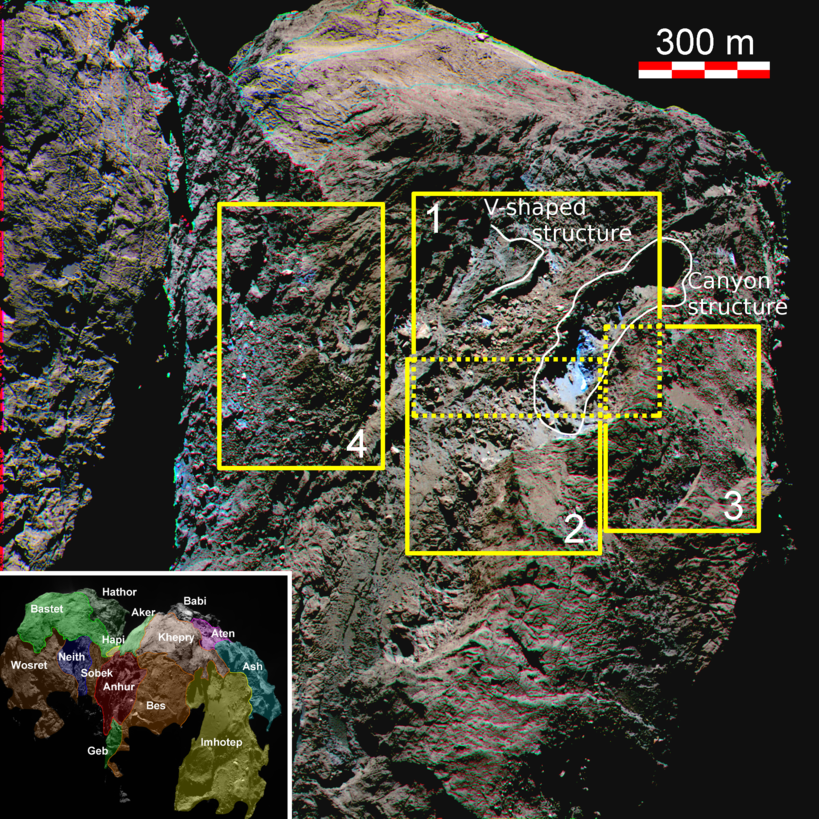}
\caption{RGB map from images obtained on 10 February 2016, UT08h14, showing the Anhur region. The yellow rectangles indicate the position of the subregions analyzed in Figs.~\ref{areaB},~\ref{areaC}, ~\ref{areaD}, and ~\ref{areaA}. The so-called V-shaped layered terrain and the canyon-like structure are also indicated. In the bottom left corner, we insert a 3D view of the southern hemisphere of 67P and overlay regional boundaries to facilitate locating the Anhur region on the nucleus.}
\label{changes}
\end{figure*}

This paper aims at investigating Anhur, a region in the southern hemisphere. The Anhur region is more fragmented than other areas on the nucleus, with fractures and patterned rough terrains (Fig.~\ref{changes}).  Anhur experiences intense thermal changes during the orbit because it is illuminated for a relatively short time interval, but at small heliocentric distances. The erosion rate that is due to the sublimation of water ice in the southern hemisphere therefore is up to 3-4 times higher than in the northern hemisphere (Keller et al., 2015, 2017). These strong thermal effects produce high erosion rates in this region. \\
Anhur is dominated by outcropping consolidated terrains that are sculpted by staircase terraces, which are interpreted as the surface expression of extended discontinuities that separate superimposed layers of consolidated material (Massironi et al. 2015; Giacomini et al. 2016; Lee et al. 2016; Penasa et al., 2017). The region also shows various types of deposits and a peculiar canyon-like structure (Fornasier et al., 2017). The consolidated terrains are crossed in places by long fracture systems superimposed on the meter-scale polygonal systems, showing pervasive thermal cracking, as described by El-Maarry et al. (2015b, 2016) for other regions of the cometary nucleus.  Numerous alcoves, pits (one showing activity events, see Fornasier et al., 2017), debris and talus deposits, boulder fields, and diamictons, located far from cliffs, are observed (El-Maarry et al., 2016; Pajola et al., 2016b; Fornasier et al., 2017). As detailed in Lee et al. (2017), the Anhur geomorphology includes both elevated terrains at higher latitude and elongated canyon-like depressions that expose the deepest regions of the large lobe of the nucleus, which presumably contains  less processed cometary material. A detailed investigation of the geomorphological properties of this region is reported in Fornasier et al. (2017, see their Fig. 2) and Lee et al. (2017).

 Moreover, the region shows local compositional heterogeneity on a scale of tenths of meters, and several bright spots are observed. Two extended bright patches (of about 1500 m$^2$ each) of exposed water ice were identified at  the end of April 2015, surviving for at least ten days (Fornasier et al., 2016, 2017). One of these (at the boundary between the Anhur and Bes regions) also included the first and so  far only detection of CO$_2$ ice, which was observed with the Visible and InfraRed Thermal Imaging Spectrometer (VIRTIS) (Filacchione et al., 2016a) in March 2015. The observations of different ices clearly point to a more pristine area where different volatiles are exposed. Anhur is also a highly active region that is the source of several jets (26 are reported by Fornasier et al., 2018, this issue) and, in particular, of the perihelion outburst. This outburst was one of the brightest activity events reported for the 67P nucleus during the Rosetta observations.
This paper focuses on the morphological and color changes observed in the Anhur region during the extended Rosetta mission in 2016 at sub-meter spatial resolution.

\section{Observations and data analysis} 

The data presented in this study were acquired with the Optical, Spectroscopic, and Infrared Remote Imaging System (OSIRIS) on board Rosetta, comprised of a Narrow Angle Camera (NAC) for nucleus surface and dust studies, and a Wide Angle Camera (WAC) for the wide-field coma investigations (Keller et al., 2007). We used radiance factor images produced by the OSIRIS standard pipeline up to level 3B, following the reduction steps  described in Tubiana et al. (2015).\\
Images were corrected for bias, flat field, geometric distortion, calibrated to absolute spectral radiance (in $W m^{-2} nm^{-1} sr^{-1}$), and finally converted into radiance factor ($I/F$, where I is the observed spectral radiance, and F is the incoming solar spectral irradiance at the heliocentric distance of the comet, divided by $\pi$), as described in Fornasier et al. (2015). \\

For the spectrophotometric analysis, the images of a given observing sequence were first coregistered using the F22 NAC filter (centered at 649.2nm) as reference and applying a python script based on the scikit-image library (Van der Walt et al., 2014) and the optical flow algorithm (Farneb\"ack, 2003). Images were then photometrically corrected applying the Lommel-Seeliger disk law ($D(i,e)$), which has been proven to give a satisfactory correction for dark surfaces (Li et al., 2015): 
\begin{equation}
D(i,e) = \frac{2\mu_{i}}{\mu_{e}+\mu_{i}}
,\end{equation}
where $\mu_{i}$ and $\mu_{e}$ are the cosine of the solar 
incidence (i) and emission (e) angles, respectively. The geometric information about the illumination and observation angles were derived using the 3D stereophotoclinometric shape model (Jorda et al., 2016). 

RGB images were generated from coregistered NAC images that were acquired with the filters centered at 882 nm, 649 nm, and 480 nm. They were optimized in false color using the STIFF code (Bertin, 2012).

The data presented here were acquired from March 2015, when the Anhur region first became visible from Rosetta, through September 2016. Details on the observing conditions are reported in Table~\ref{tab1}.

\begin{figure*}
\centering
\includegraphics[width=0.95\textwidth,angle=0]{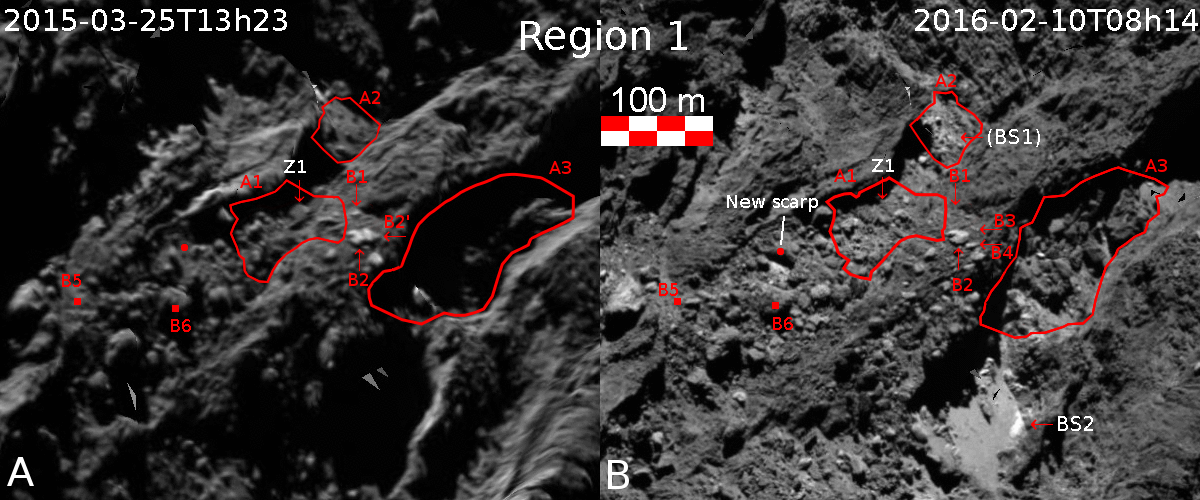}
\caption{Region 1 in Fig.~\ref{changes} in Anhur showing outcropping, consolidated terrains, and smooth surfaces (from images acquired with the orange filter on 25 March 2015 (left) and 10 February 2016 (right). The March 2015 image has a lower spatial resolution (1.7 m px$^{-1}$) than that obtained in 2016 (1 m px$^{-1}$), and it is reprojected onto the shape model to match the observing conditions of the February 2016 image. Thus the left- and right-side images  have the same spatial scale. The main surface changes are delimited by polygons  A1-A3, the fragmentation and appearance of boulders is indicated with B1-B4, and the new scarp and a changing structure are also indicated. A bright patch (BS2) is also visible in the canyon-like structure in the February 2016 image. The region A3 is in shadows in the image obtained in March 2015, but was imaged with better observing conditions in May 2015, as shown in Fig.~\ref{dustbank}. See the online movie (Area1.gif), which shows the morphological changes between these two epochs.}
\label{areaB}
\end{figure*}

 \begin{figure*}
\centering
\includegraphics[width=0.95\textwidth,angle=0]{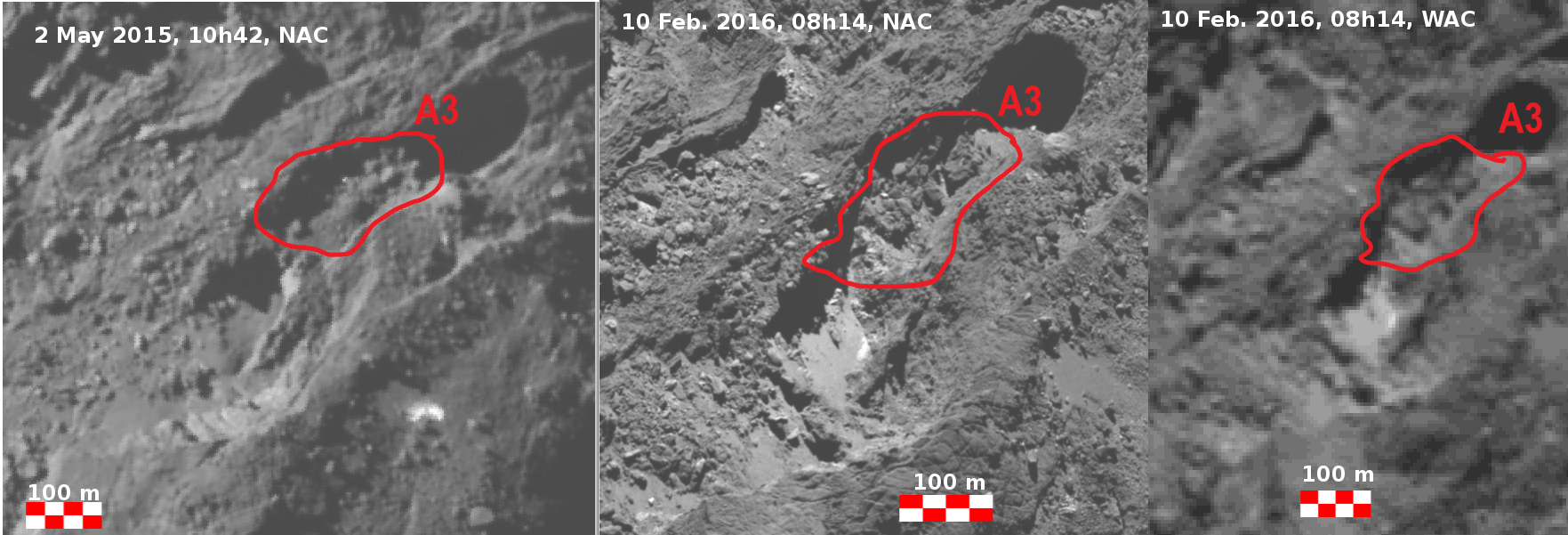}
\caption{Images from May 2015 and February 2016, showing the removal of a dust bank inside the Anhur canyon-like structure in the region delimited by the A3 polygon in Fig.~\ref{areaB}.}
\label{dustbank}
\end{figure*}

\subsection*{Measurement of mass loss}

The Anhur canyon site (Fig.~\ref{changes}) was observed at large phase angles, allowing
the use of shadow length as a proxy of height for small-scale
landmarks that are not fully defined by the latest shape models.
Shadow length has frequently been used to estimate the depth or height
of surface landmarks  in the image-resolved
studies of bodies throughout the solar system (Arthur 1974; Chappelow \& Sharpton
2002; El-Maarry et al., 2017; Hasselmann et al., 2019). The shadow length
$L_{sha}$ is measured using the generalized distance connecting the
top of a structure to the tip of its shadow, but regarding the projected
Sun direction (azimuth angle) with respect to the image frame. The generalized
distance probes the distance of two points in spherical coordinates
(i.e., imaging frame) of position $(r,\theta,\lambda)$ and $(r',\theta',\lambda')$
on the nucleus surface through the given expression:

\[
L_{sha}=
\]

\begin{equation}
\sqrt{r^{2}+r'^{2}-2r'r\left(\sin(\theta)\sin(\theta')\cos(\lambda'-\lambda)+\cos(\theta)\cos(\theta')\right)}
,\end{equation}

\noindent where $r$ and $r'$ are the spacecraft distances of two given
points obtained from the shape model, $\theta$ and $\theta'$ are
the X-axis image coordinates of the two points, in radians, referenced
to the image center, while $\lambda$ and $\lambda'$ are the same,
but for the Y-axis image coordinates. To obtain $\theta$, $\theta'$,
$\lambda$, and $\lambda'$, we multiplied their numerical position
in the image frame by the camera angular resolution.

To estimate the height $h$ of a structure using its shadow length we apply the following formula:
\begin{equation}
h=L_{sha}\cdot\tan(\pi/2-i)
,\end{equation}

where $i$ is the incidence angle estimated by the average of all
shape model facets that intercept the tops and tips of the shadow. To estimate $h$ reliably, we manually traced the shadow profile
every 5-10 pixels. For round landmarks, only a few profiles
near the summit were available. Finally, the uncertainties were calculated based
on the standard deviation from repeated measurement of the height of a given structure.

To determine the volume of the changing structures presented in Table~2, we approximated them with geometrical shapes. The missing dust cover, 
the scarp, and the associated niche were modeled with a body of quadrangular base. However, for the structure at the mouth of the canyon, which varies in height, we used a trapezoid and multiplied by its width. We then estimated the missing mass assuming the density of the bulk nucleus: $537.8\pm0.7\ kg/m^{3}$ (Patzold et al., 2016; Jorda et al., 2016; Preusker et al., 2017).
%
                          \section{Morphological changes in Anhur}
%
\begin{figure*}
\centering
\includegraphics[width=0.99\textwidth,angle=0]{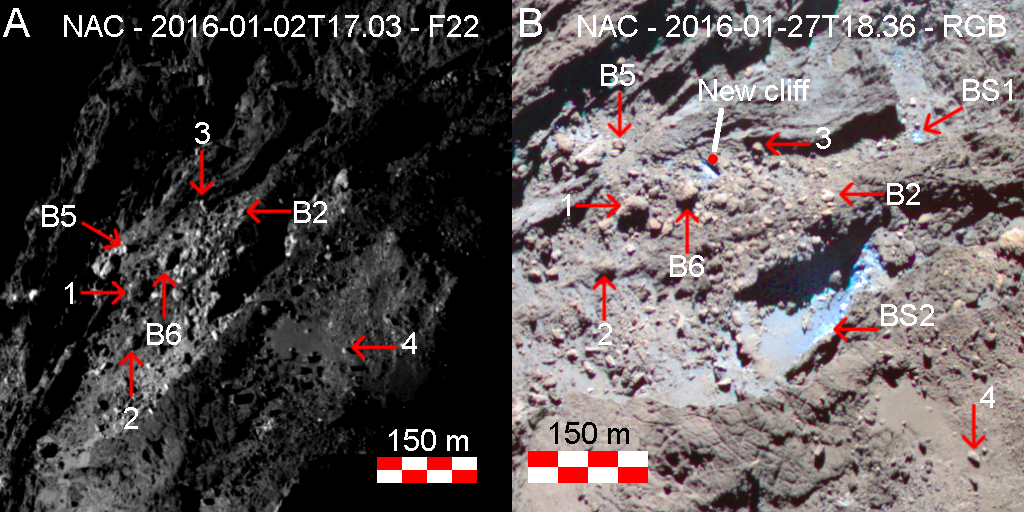}
\caption{Images acquired between 2 and 27 January 2016, showing the formation of a new scarp. Left side is an orange filter image, while on the right is a RGB image from a color sequence. In the RGB image, bright spots are visible at the base of the new cliff, as well as near the V-shaped layered terrain (BS1, inside the region named A2 in Fig.~\ref{areaB}), and inside the canyon (feature BS2). The boulders B2, B5 and B6 are the same than in Fig.~\ref{areaB}, while additional boulders for reference are indicated by numbers. }
\label{scarp}
\end{figure*}

\begin{figure*}
\centering
\includegraphics[width=0.98\textwidth,angle=0]{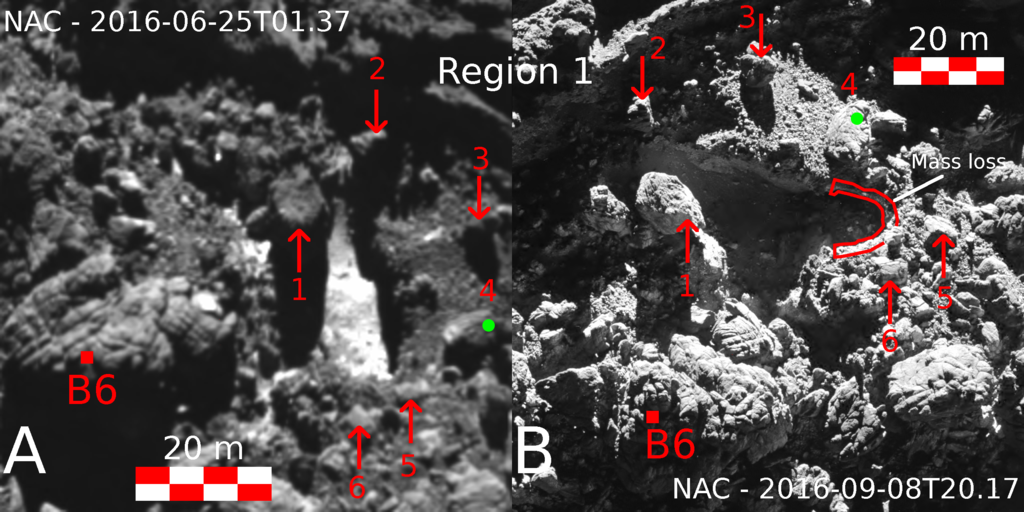}
\caption{Images acquired in June and September 2016, showing the new scarp structure at a high spatial resolution of 30 and 8 cm px$^{-1}$, respectively. The B6 boulder is the same as in Figs.~\ref{areaB} and ~\ref{scarp}. Additional boulders close to the new cliff are indicated for reference. The red U-shaped polygon indicates the mass loss in the images from June to September 2016.}
\label{scarp_hr}
\end{figure*}
\begin{figure*}
\centering
\includegraphics[width=0.8\textwidth,angle=0]{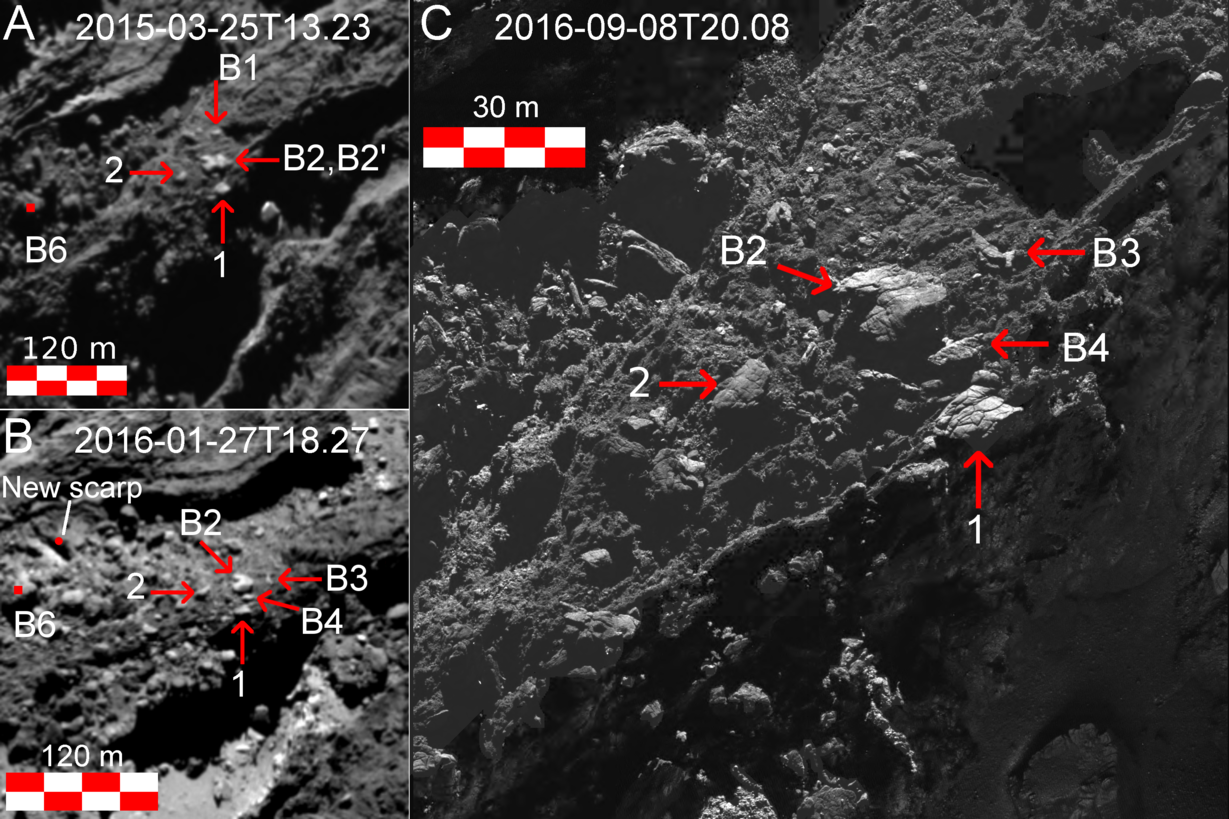}
\caption{Images from 2015 to 2016, showing the evolution of an irregular boulder of approximately 30 m length and 15 m width, which likely fragmented and generated two smaller boulders (B3 and B4) that are indicated by the red lines. The B1-B6 boulders are the same as in Fig.~\ref{areaB}. Additional boulders are indicated by numbers.}
\label{fragboul}
\end{figure*}
\begin{figure*}
\centering
\includegraphics[width=0.8\textwidth,angle=0]{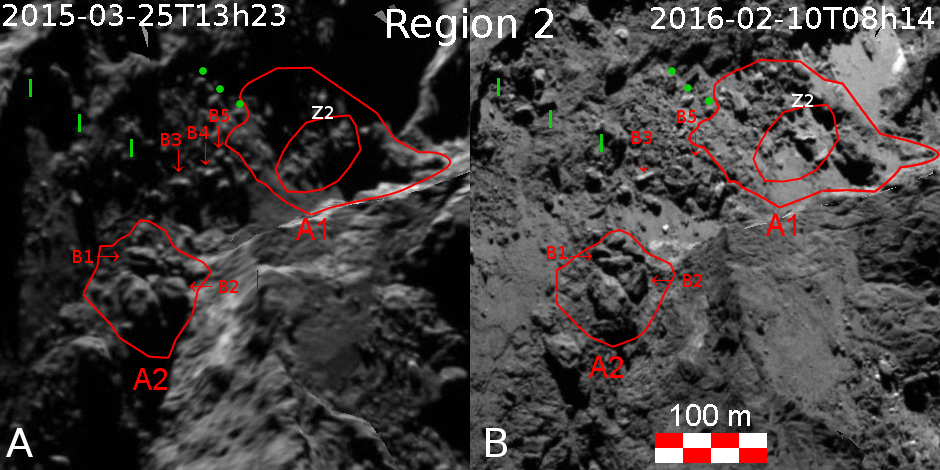}
\caption{Region 2 in Fig.~\ref{changes} in  Anhur, showing the mouth of the canyon (from images acquired with the orange filter on 25 March 2015 (left) and 10 February 2016 (right). The March 2015 image has a lower spatial resolution (1.7 m px$^{-1}$) than that obtained in 2016 (1 m px$^{-1}$), and it is reprojected onto the shape model to match the observing conditions of the February 2016 image. Thus the left- and right-side images  have the same spatial scale. Changes are indicated by the red arrows that point to different boulders (B1-B5), while the area with the most significant changes is delineated with red polygons (A1 and A2). Green points and lines indicate some reference landmarks in the images. See the online movie (Area2.gif), which shows the morphological changes between these two epochs.}
\label{areaC}
\end{figure*}
\begin{figure*}
\centering
\includegraphics[width=0.7\textwidth,angle=0]{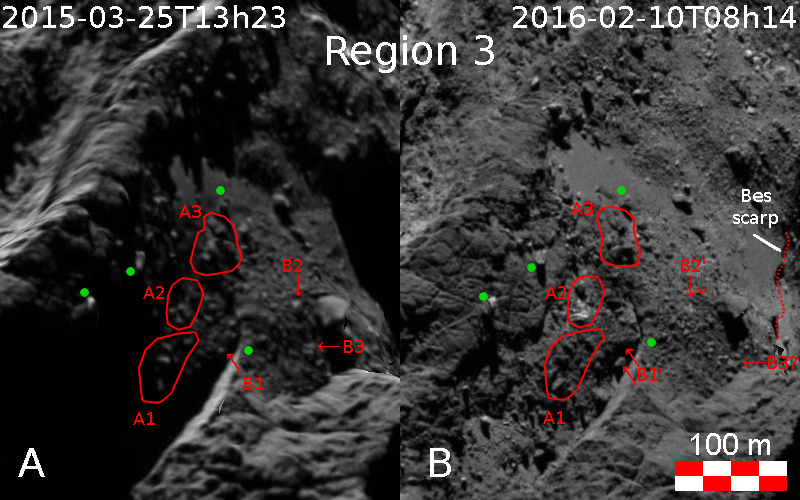}
\caption{Region  3 in Fig.~\ref{changes} (from images acquired with the orange filter on 25 March 2015 (left) and 10 February 2016 (right). The March 2015 image has a lower spatial resolution (1.7 m px$^{-1}$) than that of 2016 (1 m px$^{-1}$), and it is reprojected onto the shape model to match the observing conditions of the February 2016 image. Thus the left- and right-side images  have the same spatial scale. Features marked with red arrows point to distinct surface changes (B1-B3), and areas delineated with red polygons indicate locations where the arrangements of groups of boulders are very likely to have changed. The new scarp previously described by Fornasier et al. (2017) that is also shown in Fig.~\ref{oct2015} is indicated as well. See the online movie (Area3.gif), which shows the morphological changes between these two epochs.}
\label{areaD}
\end{figure*}

Because it is located in the southern hemisphere of comet 67P, the Anhur region only became observable from Rosetta  in March 2015, at a heliocentric distance of 2 au inbound and $\sim$ three months before the cometary equinox.  We examined more than 100 images that were obtained beginning 25 March 2015 (the date with the best spatial resolution of this region before the perihelion passage) until 8 September 2016. \\
From the comparison of images acquired at similar spatial resolution before (25 March 2015) and after (10 February 2016) perihelion passage, we detected numerous morphological changes (Figs.~\ref{areaB}--\ref{areaA}). In particular, three of these surface changes imply several million kilogram of vanishing mass through removal of dust layers and disappearance or formation of landmarks, particularly in and near the Anhur canyon (Table 2).

These changes were concentrated in four areas that we indicate by the yellow rectangles in the left part of Fig.~\ref{changes}: 1) the central outcropping consolidated terrain of Anhur; 2) the central area, close and inside a canyon like structure;  3) the area close to the boundary between Anhur and Bes;  and 4) the area close to the boundary between Anhur and Sobek.\\
The detected morphological changes fall into seven categories: \\
a) formation of scarps and cliff retreats; \\ 
b) mass loss of a relatively large structure near the mouth of the Anhur canyon; \\
c) dust cover removal inside the canyon structure;\\
d) the  displacement (rotation, shift) of boulders; \\
e) the disappearance of boulders; \\
f) the appearance of boulders; and\\
g) areal resurfacing due to one or more of these processes, which cannot be not clearly identified within the resolution or observing conditions of the images. \\
To detect  and validate these changes, we made use of the selection of images described above and of the image projection tool ShapeViewer\footnote{Available at 
http://www.comet-toolbox.com/shapeViewer.html} , which allows comparing two images that were acquired under different illumination 
conditions as if they were taken with the same viewing geometry.

A morphological change was confirmed when it was observed in multiple observations. 
When a surface change could not be clearly confirmed, it was flagged and kept as probable.\\

\subsection{Region 1} 

As shown in Figs.~\ref{changes} and ~\ref{areaB}, region 1 exhibits several changes, especially in the gravitational accumulation deposits located between the V-shaped layered material and the terrace above the canyon. The main changes are the following: dust layer removal inside the canyon-like structure, followed by the appearance of new boulders (Figs.~\ref{areaB} and ~\ref{dustbank}); a new scarp that formed in early January 2016 (Figs.~\ref{scarp} and ~\ref{scarp_hr}); two areal resurfacing events  at the foot of the V-shaped consolidated terrain (A1 and A2 in Fig.~\ref{areaB}); two new boulders (boulders B3 and B4 in Fig.~\ref{areaB} and Fig.~\ref{fragboul}) that may be related to the fragmentation of a larger boulder.


The new scarp and depressed terrain are observed at longitude [-46.76,-48.54$^{o}$] and latitude [-44.3, -43.5$^{o}$] on the western side of the canyon. This new scarp was formed after the perihelion passage.  An inspection of images acquired between March 2015 and January 2016 indicates that the scarp formed between 2 and 27 January 2016 (Fig.~\ref{scarp}). 

OSIRIS acquired high-resolution images (spatial scale between 8 and 30 cm px$^{-1}$) of the scarp close to the end of the Rosetta mission (Fig.~\ref{scarp_hr}), allowing us to investigate this structure in more detail. The scarp has a height of about 10$\pm$2 m, and the floor at the base of the scarp has an estimated surface of at least 320 m$^2$, based on the visible and illuminated surface. Considering the illuminated and shaded surface (Fig.~\ref{scarp_hr}) projected on the shape model, we estimate an upper limit of the surface of 570 $m^2$ based on June 2016 observations. The height of the cliff was estimated using two methods: a) direct measurements based on the 3D shape model (shape 8, QMAPS v1.0), and b) measurements derived from the shadow length, as detailed in section 2.   The  total sublimated mass is estimated to exceed 1.7$\times$10$^6$ kg (Table~2). New boulders and displaced terrain are visible there and nearby, as is exposed water ice, as discussed in the following section.

At the base of the scarp, the illuminated surface that became visible in January 2016 is smooth and has very few scattered boulders. Vanishing masses that left almost no large scattered boulders were also observed in the Khonsu region (Hasselmann et al., 2019). This lack of boulders seems to be connected to a very fine-grained dusty material on the sub-surface terrain that contrasts with the ice-poor crust that covers the uppermost surface.\\
 The last image acquired in September 2016 clearly shows the base of the new scarp and indicates the further crumbling of one edge of the scarp, with some mass waste on its bottom right side,  marked by the red U-shaped polygon in Fig.~\ref{scarp_hr}. Compared to the June 2016 observations, this last image shows that the cliff is closer to the boulder called  number 4 in Fig.~\ref{scarp_hr}, indicating that the scarp retreated at least 5 m in two months. Throughout the perceived retreat of the scarp, talus at the bottom left side of the scarp further suggests the ongoing evolution of this structure.  

In the A1 area (Fig.~\ref{areaB}), the most obvious changes concern a structure that is indicated by a red arrow and labeled "Z1". This structure clearly changed, and nearby, at least a few new boulders are present that were not visible in the 2015 images.  \\
Between the A1 area and the canyon, some new boulders (B3 and B4 on the right side of Fig.~\ref{areaB}) appear close to an irregular and larger boulder that is approximately 30 m long and 15 m wide (boulder called B2 and B2' in Figs.~\ref{areaB}, and~\ref{fragboul}). This boulder appeared to be smaller in the 2016 observations. Presumably, these new boulders result from a partial fragmentation of the larger boulder. Alternatively, the B2' part of the large boulder may have been covered by a dust deposit that fell from the nearby consolidated terrain and eventually also brought the two new boulders that were observed later. However, we are unable to evaluate the evolution of the depth of the dust cover in this region with the spatial resolution that is available. Another boulder, called B1 in Fig.~\ref{areaB}, was present in March 2015 images but not in those of 2016. It probably fell to the bottom of the terrace.

The A2 area (Fig.~\ref{areaB}) presents a completely different appearance in 2016 than in 2015. The number and positions of boulders clearly changed, and the surface appears rougher in 2016, which could be due to a thinning of the dust mantle in this region. Furthermore, we note that in this area, bright patches of water ice were observed in January 2016 and later in June 2016 images (Fig.~\ref{scarp}, feature called BS1,  and Fig.~\ref{RGBjune2016}). 

\begin{table*}
\begin{center}
\label{massest}
\caption{Characteristics of the largest mass loss features observed in Anhur. Z2 is the structure observed in 2015  and reported in Fig.~\ref{areaC}.$^{*}$ : we also report the estimation from Fornasier et al. (2017) for the new scarp in Bes region, nearby the Anhur-Bes boundary.}
\begin{tabular}{ccccccc}
\hline 
Landmark & OSIRIS F22 image & $\Delta$(km) & m~px$^{-1}$ & $h$ (m) & $i$ (deg.)  & $L_{sha}$ (m)\tabularnewline
\hline 
\hline 
Dust bank (Region 1, Fig.~\ref{dustbank})& NAC\_2016-02-10T08.14.06.243Z & 50.06 & 0.937 & $14\pm2$ & $42\pm6$ & $12.9\pm1.5$\tabularnewline
\hline  
New scarp (Region 1, Fig.~\ref{areaB}) & NAC\_2016-02-10T08.14.06.243Z & - & - & $10\pm2$ & $52\pm12$ & -\tabularnewline
\hline 
Z2 (Region 2, Fig.~\ref{areaC})  & NAC\_2015-03-25T13.58.49.354Z & 87.37 & 1.64 & $8.6-26.8$ & $63\pm18$ \tabularnewline
\hline
New cliff (Region 4, Fig.~\ref{areaA}, A3) & NAC\_2016-02-10T08.14.06.243Z & 49.91 & 0.934 & $9\pm2$ & $45\pm11$ & $9\pm2$\tabularnewline
\hline 
Cliff retreat (Region 4, Fig.~\ref{areaA})& NAC\_2015-03-25T13.23.48.562Z & 87.67 & 1.64 & $4\pm2$ & $76\pm1$ & $14\pm3$\tabularnewline
 &  &  &  &  &  & \tabularnewline
\hline 
\end{tabular}
\par
\begin{centering}
\begin{tabular}{cccccccc}
\hline 
Landmark &  length (m) & width (m) & Area (m$^2$) & Volume (m$^3$) & Mass (kg) & \tabularnewline
\hline 
\hline 
Dust bank (Region 1, Fig.~\ref{dustbank})& $116\pm2$ & $(53-73)\pm2$ & 949 & 13145 & $7.0\cdot10^{6}$ & \tabularnewline
\hline 
New scarp (Region 1, Fig.~\ref{areaB})& $26\pm2$ & $12\pm2$ & 327.5 & 3144 & $1.7\cdot10^{6}$ & \tabularnewline \hline 
Z2 (Region 2, Fig.~\ref{areaC}, A3)&  $92\pm4$ & $(17-25)\pm4$ &  &  &   $1.9\cdot10^{7}$\tabularnewline
\hline 
New scarp in Bes$^{*}$ (Region 3, Figs.~\ref{areaD}, and ~\ref{oct2015}) & 140 & & $\sim$5000 & 50000 & $2.6\cdot10^{7}$& \tabularnewline \hline 
New cliff (Region 4, Fig.~\ref{areaA}, A3) & $(36/44)\pm2$ & $(9.1/24.3)\pm2$ & 215/437 & 1920/3840 & $1-2\cdot10^{6}$ & \tabularnewline
\hline 
Cliff retreat (Region 4, Fig.~\ref{areaA}, A2) & $9\pm4$ & $15\pm4$ & 43 & 153 & $8.2\cdot10^{4}$ & \tabularnewline
\hline 
\hline 
\end{tabular}
\par\end{centering}
\end{center}
\end{table*}

The most extensive mass loss occurred inside the canyon (Figs.~\ref{areaB} and ~\ref{dustbank}), located within, or close to the estimated source of the so-called perihelion outburst (see Fig. 11 in Fornasier et al., 2018, this issue) that took place on 12 August 2015, one day before perihelion passage. A rough 45 m long terrace (lat=-45$^o$, lon=60$^o$) is revealed in images of February 2016, where in images of May 2015, only a very smooth terrain was observed (Fig.~\ref{dustbank}).  New boulders are clearly visible in the February 2016 images at different spatial resolutions acquired with the NAC and WAC cameras. The new structures are still visible in the lowest resolution images acquired with the WAC camera, indicating that the morphological changes are real and not related to the different spatial resolutions between the pre- and post-perihelion images.
Locally, the site lowered by $\sim 14\pm2$ meters, as estimated
by the height of the new structure. The area affected by the changes is difficult to estimate but considering the surface that appeared to be covered by dust in the May 2015 image (the central part of the canyon in Fig.~\ref{dustbank}), we find that at least a mass of roughly $7\times 10^{6}$ kg (Table~2) has vanished. A similar transformation process in smooth terrains and the formation of a 14 m deep cavity has also been reported in the southern equatorial Khonsu region (Hasselmann et al., 2019). 

\subsection{Region 2} 

Moving to the south, in  region 2 (Figs.~\ref{changes} and ~\ref{areaC}), near the canyon mouth (feature called Z2 inside the A1 polygon in Fig.~\ref{areaC}), we find a landmark that has completely disappeared in February 2016 and only left a remaining block. This structure consists of consolidated material or boulders over a smooth terrain. The measurement of its shadow length on the image NAC 2015-03-25 UTC 13:58:49 results in a steep height slope that extends from 8.6 to 26.8 meters in height. From the estimated length of $92\pm4$ meters and width of $(17-25)\pm4$ meters, we calculate a total mass loss of about $1.9\times10^{7}$ kg (Table~2).

In Fig.~\ref{areaC}, the A2 polygon points to the bottom of a scarp where two boulders (B1 and B2) have moved. Nearby, the regions in between A1 and A2 also show boulder displacements. Unfortunately, OSIRIS did not acquire images with observing conditions that are good enough to quantify the changes that occurred in this area (a possible cliff retreat or mass loss that perhaps also affected the positions of several boulders). For example, the boulder B4 is no longer visible in February 2016 images and may either have moved away from this region or been shifted toward the neighboring smooth material.

\begin{figure}
\centering
\includegraphics[width=0.45\textwidth,angle=0]{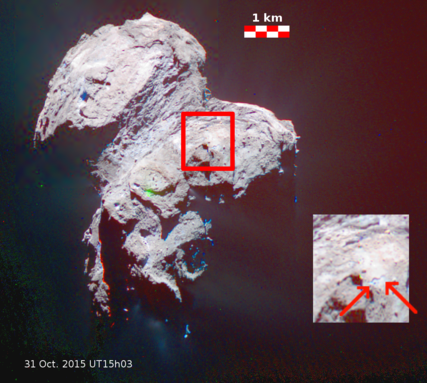}
\caption{RGB images (from the images acquired with filters centered at 882, 649, and 480 nm) from 31 October 2015 UT15h07 observations. The rectangle indicates the new scarp, described first in Fornasier et al. (2017).}
\label{oct2015}
\end{figure}

\subsection{Region 3} 

In region 3 (Figs~\ref{changes} and ~\ref{areaD}) we observe changes in three boulders. Boulder B1, emerging from the shadows, is shifted from its original position of about 10 m, while boulder B2 disappeared; a bright feature appeared  near its original position. Boulder B3  is not located in the same position in the February 2016 images. In the 2016 image, a boulder with a similar size as boulder B3 in the 2015 image is visible. Thus, our interpretation is that the boulder presumably shifted close to the base of the nearby cliff.  We also note that the surrounding smaller boulders changed position.  

The A1--A3 areas in Fig.~\ref{areaD} mark the locations where the boulder positions have likely changed between the two epochs. The more notable changes are within region A2, where a group of boulders seems to have been displaced between the two epochs, followed by the appearance of a brighter surface in the 2016 images. However, because of the illumination conditions in the 2015 images, which have extensive shadows on the examined areas, we are unable to unequivocally quantify the changes. 

Fig.~\ref{areaD} also indicates a new scarp that was previously described by Fornasier et al. (2017, see their Fig. 13). This new scarp is 140 m long and about 10 m high. It was observed at the boundary of the Anhur and Bes regions, located in and nearby an extended bright water-ice-rich patch of about 1600 m$^2$ (called patch B in Fornasier et al., 2016, 2017), which appeared at the end of April 2015 and survived for at least ten days. The formation time of this new scarp was not well constrained. Fornasier et al. (2017) provided a temporal range for its formation between 1 August 2015 and 10 December 2015. After a  careful inspection of the images in the OSIRIS archive, we found an image on 31 October 2015 where the scarp was still visible (Fig.~\ref{oct2015}). This allowed us to better constrain the scarp formation time to near  perihelion or a few weeks after. In Fig.~\ref{oct2015}, a few jets are also visible on the nucleus; they are fully analyzed in Fornasier et al. (2018). The formation of this scarp implies a maximum volume loss of about 50000 m$^3$ (Fornasier et al., 2017). When we assume for the loss material the same bulk density as for the comet, about 2.6$\times$10$^{7}$ kg of cometary material sublimated during the scarp formation, thus exposing the deepest layers.  \\
Interestingly, high-resolution images of this scarp acquired on May 2016 (at a spatial scale of 20 cm px$^{-1}$) indicate spectrally blue material and bright patches at the foot of the scarp. Fornasier et al. (2017) estimated a water-ice abundance of 17$\pm$2\% for the brightest spot observed there in 2016 based on a linear mixing model of water ice (30 $\mu$m grain size) and the
dark terrain of the comet.

\noindent

\subsection{Region 4} 

\begin{figure*}
\centering
\includegraphics[width=0.94\textwidth,angle=0]{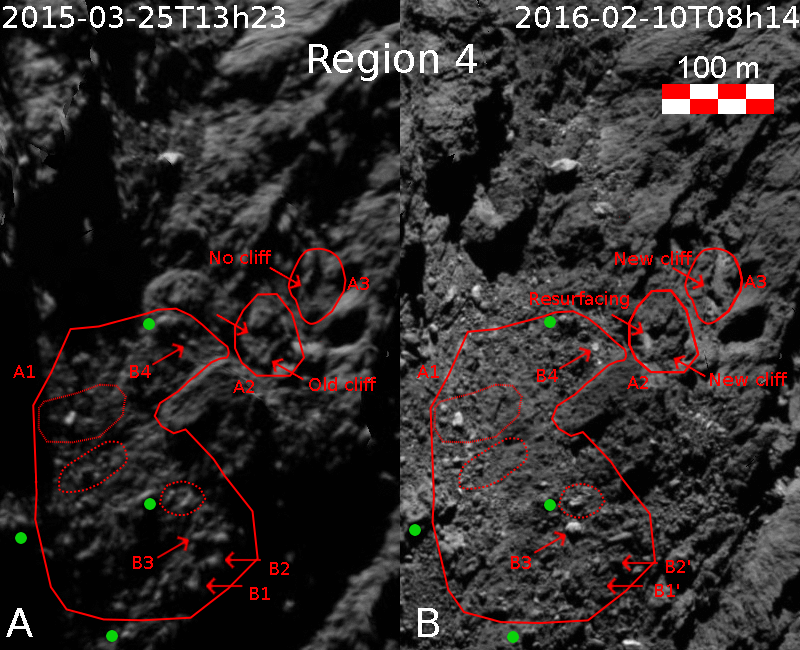}
\caption{Region 4 in Fig.~\ref{changes} in  Anhur  (from images acquired with the orange filter on 25 March 2015 (left) and 10 February 2016 (right). The March 2015 image has a lower spatial resolution (1.7 m px$^{-1}$) than that in 2016 (1 m px$^{-1}$), and it is reprojected onto the shape model to match the observing conditions of the February 2016 image. Thus the left- and right-side images  have the same spatial scale. Changes are observed  within the A1-A3 polygons, notably a cliff retreat in A2 area and a new cliff in area A3, as well as in the location and appearance of boulders B1-B4. See the online movie (Area4.gif), which shows the morphological changes between these two epochs.}
\label{areaA}
\end{figure*}

Finally, in region 4 (Figs.~\ref{changes} and ~\ref{areaA}), located near the boundary between Anhur and Sobek, we have identified three areal resurfacings, the formation of a new cliff, and a cliff retreat (Fig.~\ref{areaA}). In polygons A1 and A2, several locations that are indicated by dotted areas in A1 and red arrows in A2 appear to have experienced some resurfacing in terms of boulder location and surface texture. Additionally, the boulders indicated by the  arrows in area A1 have shifted or moved away. In area A2, we identified a cliff retreat over about 10 m. The mass loss produced by this resurfacing is minor (80000 kg, Table 2) compared to other changes we investigated here. The estimated mass loss produced by the new cliff, $\sim$ 9m in height, observed within area A3 (Fig.~\ref{areaA}), is more relevant : it is on the order of 1-2 $\cdot 10^{6}$ kg. The factor 2 in the mass loss estimate is related to the difficulties of evaluating the original extension of the structure. Several other minor changes (boulder displacements and roughness changes) are present, but cannot be quantifed because the spatial resolution of pre-perihelion images is limited.

\section{Activity in Anhur: perihelion outburst}

The so-called perihelion outburst that took place on 12 August 2015 at 17:20 (number 14 in Table 1 in Vincent et al., 2016a, number 8 in Table A.1 in Fornasier et al., 2018, this issue) was one of the brightest events detected by Rosetta.
Vincent et al. (2016a) and Fornasier et al. (2018, see their Fig. 11) have previously shown that this jet originated in Anhur, more precisely, from the canyon structure (lon=59.9$\pm$5.8$^{o}$, lat=-52$\pm$12$^{o}$). \\
The mass of this event was previously estimated in Lin et al. (2017). They found a lower limit of 5860 kg for the collimated component of the jet, using a grain-size distribution with a differential power-law index of -3.7 (from 1 $\mu$m to 1 mm) and an albedo of 0.068 (Fornasier
et al., 2015). We aim at giving a better mass-loss estimation using the latest dust size-distribution power laws and coma phase curves (Bertini et al., 2017) derived for comet 67P.

The outburst dust mass in the 649 nm (orange filter) image is estimated in three
steps. (i) The pixel-wise filling factor is the radiance
factor normalized by the average dust phase curve for the event, approximated by that of the inner coma (Bertini et al., 2017). (ii) The mass is obtained by multiplying the filling factor by the 
area covered by the outburst, by the average grain density (795 $kg/m^{3}$, Fulle et al.,
2016a), and by the grain size distribution (Agarwal et al., 2017) in the mathematical
formulation reported in Fornasier et al. (2018, this issue), Hasselmann
et al. (2019), or Rinaldi et al. (2019). For
the perihelion jet we considered a single-scattering albedo
representing a dust that is dominated by dark carbon-rich grains (0.045, Hasselmann
et al., 2019). Because brighter material such as silicates and ices might be present in the outburst, the ejected mass estimate is an upper limit. In this step we convert the radiance factor into mass per pixel. (iii) We then selected an area that enveloped all components of the event and another so-called control area (Fig.~A.1), which has the same number of pixels as the first area, to evaluate the contribution of the quiescent coma, which is later subtracted from the total
mass. Hassselmann et al. (2019) have estimated an uncertainty
of 15-20\% in mass that is only due to fluctuations in the quiescent coma
contribution.

By applying these steps to the 12 August 2015 UT17h20 image acquired with the orange filter, when the outburst was at its maximum level, we estimate a dust
mass of $(1.8\pm0.3)\cdot10^{6}$ kg, which is comparable to the
total mass of the dust cover lost (see section 3.1) in the central
part of Anhur canyon. At UT 17:35, the outburst is still observed to
carry a mass of $(1.0\pm0.3)\cdot10^{6}$ kg, roughly half the
mass ejected 15 minutes before. Half an hour later, at UT 18:05, the jet
is no longer detected, and the inner coma shows only dust filaments. Before the outburst, OSIRIS acquired an image at UT 17h05, showing a small activity. Thus we may constrain the event duration to between 15 and 60 minutes. Consequently,
the dust production rate ranges between 289 and 1990 kg/s. The maximum
value is consistent with the peak in the coma dust production rate,  
which is estimated to be up to 4500 kg/s near perihelion for a steady dust-loss rate (Fulle et al., 2019) . 

\section{Evidence of frost and exposed water ice}

\begin{figure*}
\centering
\includegraphics[width=0.98\textwidth,angle=0]{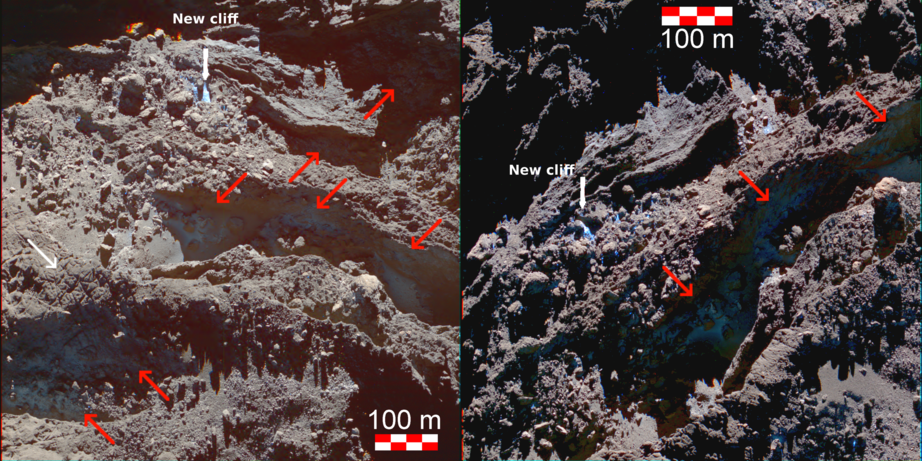}
\caption{RGB composite (from the images acquired with filters centered at 882, 649, and 480 nm) of the Anhur region from observations acquired on 25 June 2016 at UT01h37 (left) and 11h50 (right). Red arrows point to the detection of frost inside the canyon-like structure and in other shadowed regions. The white arrow indicates orthogonal fractures. Several exposures of water ice are also detected, notably at the base of a new scarp.}
\label{RGBjune2016}
\end{figure*}
\begin{figure*}
\centering
\includegraphics[width=0.98\textwidth,angle=0]{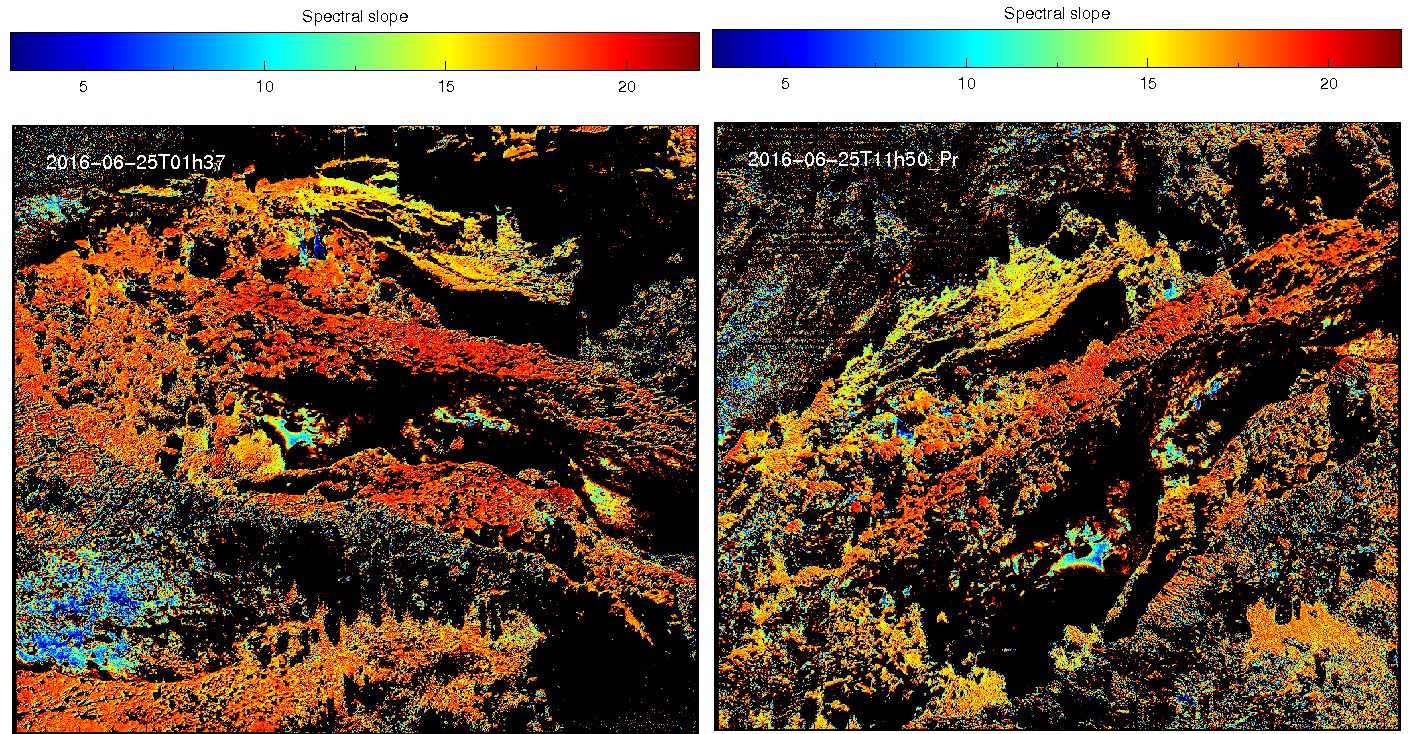}
\caption{Spectral slope evaluated in the 535-882 nm range for the 25 June 2016 observations (left: 01h37; right: 11h37).}
\label{slopesjune}
\end{figure*}
\begin{figure*}
\centering
\includegraphics[width=0.9\textwidth,angle=0]{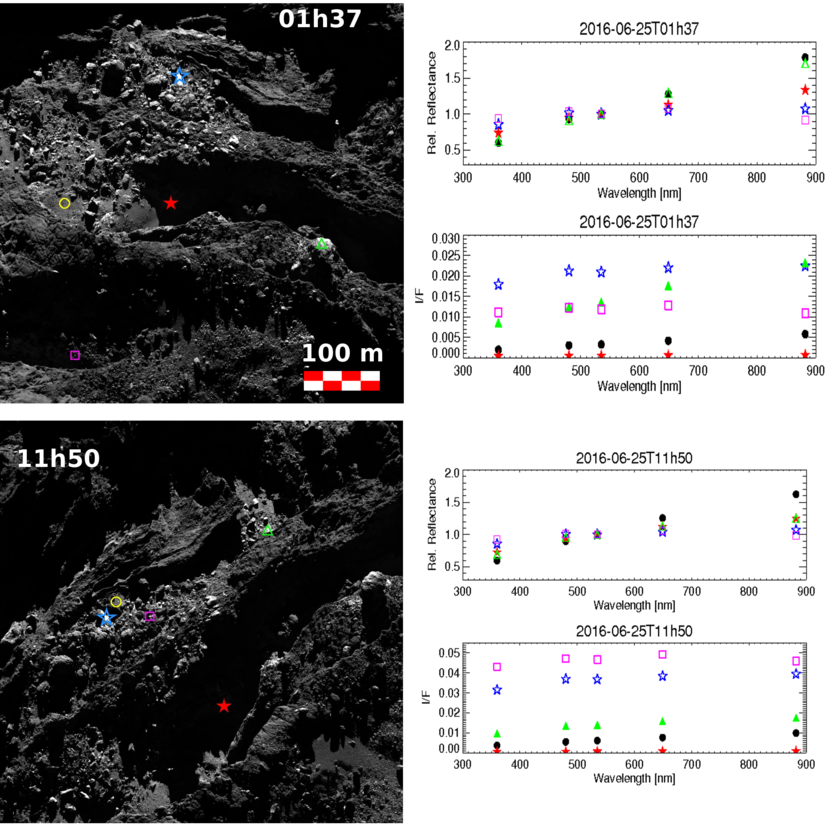}
\caption{Analysis of the 25 June 2016 data (top: 01h37; bottom: 11h50). Location of 5 ROIs (left), and their relative reflectance and radiance factor, at phase angle = 87$^o$ (right). The dark terrain is represented by a circle in yellow for clarity in the black-and-white images on the left side and in black in the plots (right).}
\label{juneROI}
\end{figure*}
\begin{figure*}
\centering
\includegraphics[width=0.45\textwidth,angle=0]{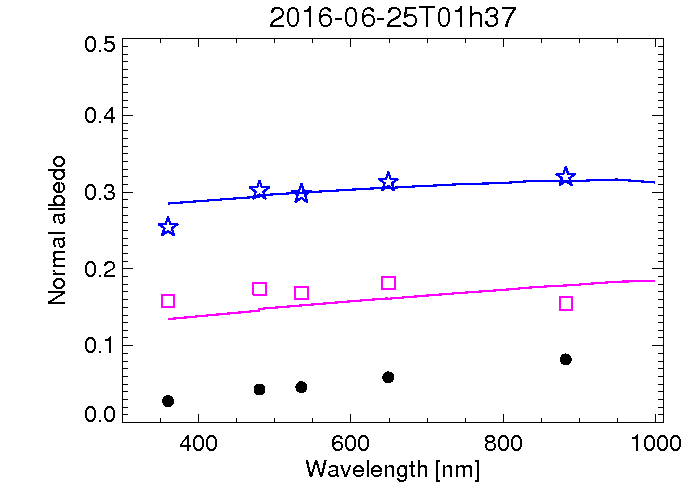}
\includegraphics[width=0.45\textwidth,angle=0]{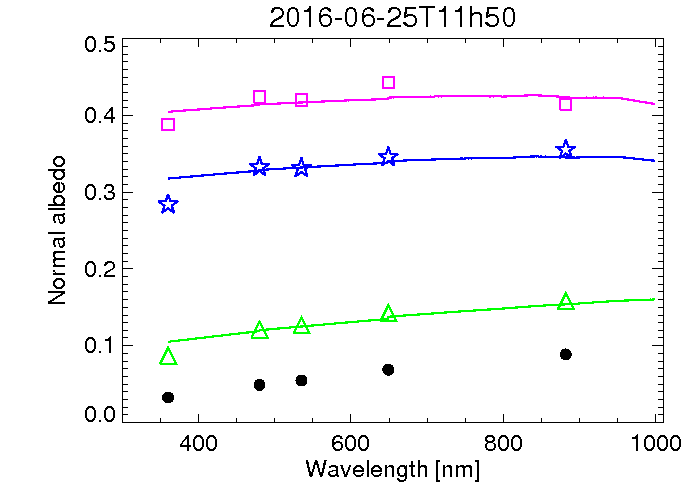}
\caption{Linear mixing models of the cometary dark terrain (represented by the dark circle) with water ice for the brightest features observed on 25 June 2016 at UT01h37 (left) and 11h50 (right). Colors and symbols are the same as for the ROIs represented in Fig.~\ref{juneROI}.  The estimated water-ice content is reported in Table~\ref{tab_model}.}
\label{model_june2016}
\end{figure*}
\begin{table*}
         \begin{center} 
         \caption{Water-ice content estimates for the bright and spectrally flat features observed in 25 June 2016 images. Symbols refer to the ROIs represented in Fig.~\ref{juneROI}. The water-ice abundance was estimated using a linear mixing model of water ice and cometary dark terrain. For completeness, we also report the estimate of the water-ice abundance at the base of the new scarp from a color sequence acquired on 27 January 2016, shortly after the new structure formed. These values should be compared with those indicated by the blue star, which correspond to the same location at the base of the new scarp. For the frost pointed out by the red star, we only report the spectral slope because the normal albedo cannot be estimated as the frost lies in shadow. Errors in the spectral slope are on the order of 0.5\%/100nm.}
         \label{tab_model}
        \begin{tabular}{|llll|} \hline
time          &  ROI & Spectral slope & Ice content (\%)  \\ 
              &      & [\%/100nm]     &                        \\ \hline

2016-01-27T18h36 & new scarp & 10.3 & 11.3 \\ \hline
2016-06-25T01h37 & circle & 20.3 & 0, DT     \\
2016-06-25T01h37 & red star & 9.7   &  --  \\
2016-06-25T01h37 & blue star (new scarp) &   2.1   & 26.5  \\
2016-06-25T01h37 & green triangle  & 20.3   &  -- \\
2016-06-25T01h37 & magenta square  & -2.4    & 11.0 \\   \hline
2016-06-25T11h50 & circle & 17.9 & 0, DT     \\
2016-06-25T11h50 & red star &     7.2   &  --\\
2016-06-25T11h50 & blue star (new scarp) &     2.1   & 29.5  \\
2016-06-25T11h50 & green triangle  &    7.15   & 7.5 \\
2016-06-25T11h50 & magenta square  & -0.4   &  38.0 \\   \hline

        \end{tabular}
\end{center}
 \end{table*}

Anhur is found to have heterogeneities in composition at scales of decimeters and exposures of ices. Before the high-resolution images acquired during the Rosetta extended mission in 2016,  evidence of exposed volatiles was reported in Anhur based on a number of different observations. \\
i) The first detection of CO$_2$ ice at the boundary between the Anhur and Bes regions (located at lon=66.06$^{o}$ and lat=-54.56$^{o}$) on 21-22 March 2015, when the southern hemisphere started to be illuminated by the Sun after its long cometary winter. This 80$\times$60 m$^2$ area was estimated to contain about 57 kg of carbon dioxide, corresponding to a 9 cm thick layer (Filacchione et al., 2016a). ii) The detection of two bright patches of exposed water ice, the largest observed by Rosetta (more than 1500 m$^2$ each), on 27 April 2015, which survived for at least 10 days: one located in the same position as the CO$_2$ ice detection, and the other centered at  lon = 76.45 $^{\circ}$ and lat = -54.15$^{\circ}$, in the Bes region, but very close to the boundary with Anhur. Fornasier et al. (2016, 2017) estimated that these patches contain 20–30\% of water ice mixed with darker material, forming a layer of solid ice that is up to 30 cm thick. iii) The observations of several smaller and localized bright spots with a flat spectral behavior, indicating that they are enriched in water ice; these bright spots were reported in Fornasier et al. (2017) and are located close to shadows and/or in the pit deposits, mostly inside the canyon-like structure.

Here we report the spectral analysis of the Anhur region from two high-resolution ($\sim$ 30 cm px$^{-1}$) sets of images taken on 25 June 2016. The two RGB images that were acquired about ten hours apart (Fig.~\ref{RGBjune2016}) clearly show the exposure of water ice on illuminated areas and frost in shadowed areas, especially within the canyon-like structure. This confirms that Anhur is relatively rich in volatiles compared to other nucleus regions (Fig.~\ref{RGBjune2016}). \\
For several regions of interest (ROI), notably bright patches, we computed the relative spectrophotometry and the radiance by integrating them over a box of 3$\times$3 pixels. The associated spectral slopes, computed in the 882-535 nm wavelength range and normalized at the green filter centered at 535 nm,  are reported in Fig.~\ref{slopesjune}. The spectral slopes evaluated in June 2016 at large phase angle (about 87$^o$) have relatively high values (average slope $\sim$ 18 \%/100 nm) on the terraces and consolidated materials, but locally, their values decrease in some large smooth areas. In particular, the spectral slope is close to zero in very localized and bright areas (Table~\ref{tab_model}), such as the niche at the base of the new scarp; this indicates exposure of water ice. \\
We thus define relatively blue regions as those with a spectral slope value between 3 and 10 \%/100nm at phase angle 87$^o$, and whose  reflectance does not exceed three times that of the cometary dark terrain.  The brightest regions, which have higher abundances of water ice, are defined as having an absolute reflectance higher than at least three times that of the cometary dark terrain, and a spectral slope lower than 3\%/100 nm, as is the case at the base of the new scarp.  Regions that are both bright and neutral to negative spectral slope according to the criteria listed above, indicating exposure of fresh water ice, cover a total surface area of $\sim$ 39 m$^2$ and 64 m$^2$ for the 01h37 and 11h50 observations, respectively. \\
We report the relative reflectance and I/F at phase 87$^o$ of 5 ROIs in each set of observations in Fig.~\ref{juneROI}. Tiny bright patches (indicated by the magenta squares in Fig.~\ref{juneROI}) and the base of the new scarp (blue star in Fig.~\ref{juneROI}) are 4-6 times brighter than the cometary dark terrain and show a flat spectrum; this is consistent  with the presence of water ice.  

To estimate the water-ice content of the bright features, we first evaluated the normal albedo from the photometrically corrected images and the Hapke model parameters determined by Fornasier et al. (2015, see their Table 4) from resolved photometry of the comet in the orange filter centered at 649 nm. On the photometrically corrected spectrophotometry of a given ROI, we then performed a simple linear mixing model with two components: the cometary dark terrain (represented by the circle in Fig.~\ref{juneROI}) and water ice,
\begin{equation}
R =  p \times R_{ice} + (1-p) \times R_{DT}
 ,\end{equation}
where $R$ is the reflectance of the bright patches, R$_{ice}$ and R$_{DT}$ are the reflectance of the water ice and of the cometary dark terrain, respectively,  and p is the relative surface fraction of water ice. The water-ice spectrum was derived from the synthetic reflectance from Hapke modeling starting from optical constants published in Warren and Brandt (2008) and adopting a grain size of 30 $\mu$m, as is typical  for ice grains on cometary nuclei (Sunshine et al., 2006; Capaccioni et al., 2015; Filacchione et al., 2016b). 
Areal (linear) mixture models were used because reliable and relevant optical constants for the dark material needed to run more complex scattering models are not available. The results of this model for bright features are presented in Fig.~\ref{model_june2016} and Table~\ref{tab_model}. \\

The high-resolution images shown point to very localized exposures of water ice, with higher abundances (Table~\ref{tab_model}) than those estimated in other regions of the comet from VIRTIS observations up to April 2015 (i.e., a few percent, de Sanctis et al., 2015; Filacchione et al., 2016b; Pommerol et al., 2015; Barucci et al. 2016; Oklay et al., 2016), but consistent with higher amounts ($>$ 20\%) determined from OSIRIS observations in localized areas in the Anhur, Bes, Khonsu, and Imhotep regions (Fornasier et al., 2016; Deshapriya et al., 2016; Oklay et al., 2017; Hasselmann et al., 2019), and at the Aswan site (Pajola et al., 2017). 

Interestingly, the new scarp (Figs~\ref{scarp} and ~\ref{RGBjune2016}) shows exposure of water ice since its formation in January 2016.  Following the same method as described above, we estimate a water-ice content of 11\% from the January 2016 images, when the base of the newly formed scarp was not as bright as in June 2016 and had a steeper spectral slope (10\%/100 nm, see Table~\ref{tab_model}). This points to a progressive enrichment in the exposed water ice in this area from January to June 2016 and to the persistence of volatiles for at least six months. A similar long survival of bright patches was found elsewhere on the comet, notably for a bright spot in the Khonsu region (Deshapriya et al., 2016), blue and bright regions in Imhotep (Oklay et al., 2017), and for the Aswan cliff collapse (Pajola et al., 2017).
\begin{figure*}
\centering
\includegraphics[width=0.9\textwidth,angle=0]{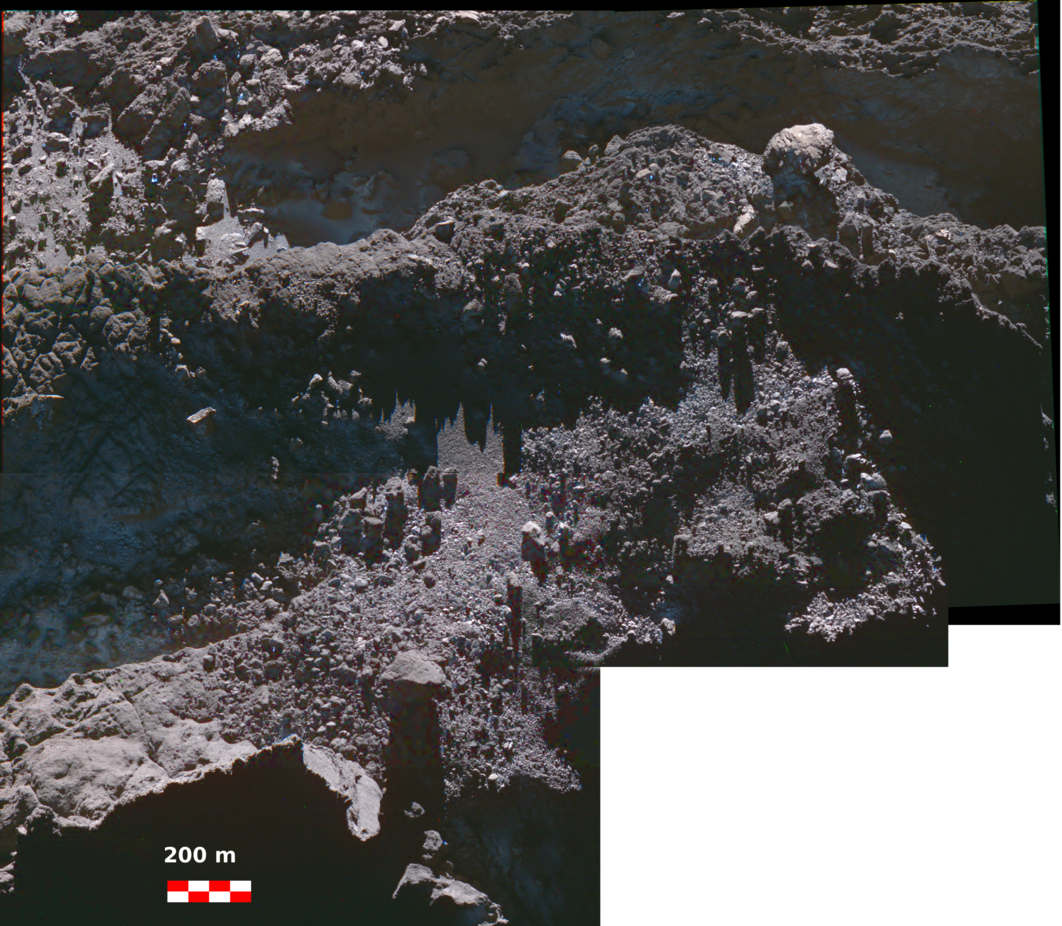}
\caption{Composite RGB images (from the images acquired with filters centered at 882, 649, and 480 nm) of four individual observations acquired on 30 July 2016, between UT 05h02 and 05h12, and at a spatial scale of 17 cm px$^{-1}$. Several tiny bright patches and extended areas covered by frost are visible close to or within shadowed regions.}
\label{30july05}
\end{figure*}

Frost and water ice in the form of tiny bright spots near shadowed regions was not unique to the June 2016 observations but was repeatedly observed later, in July 2016. We report an example for the southern part of Anhur in Fig.~\ref{30july05}, which is an RGB composite from four observations carried out between 05h02 and 05h12 on 30 July 2016, at a spatial scale of 17 cm px$^{-1}$. Several tiny bright patches, especially close to shadowed regions near boulders, are observed, as is frost inside the canyon structure and in other shadowed areas.\\

\begin{figure*}
\centering
\includegraphics[width=0.99\textwidth,angle=0]{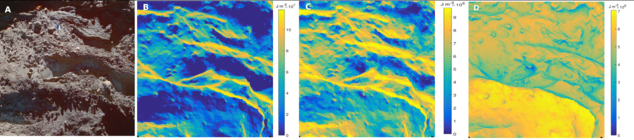}
\caption{Modeled accumulated insolation (in J m$^{-2}$) in the Anhur region as viewed on 25 June 2016 UT01h37 (A)  during time periods of  one month (panel B, May 25 --June 25, 2016), six months (panel C, January 1 --June 25, 2016), and  one year (panel D, June 25, 2015 – June 25, 2016).}
\label{insolation}
\end{figure*}

The canyon-like structure hosts frost because it is often in shadow, resulting in a low solar insolation. To prove this, we investigated the history of illumination in the Anhur region. Using the shape model of the comet with resolution of $\sim$ 10 m from Preusker et al. (2017), together with ephemerides and rotational status of 67P from SPICE kernels (Acton, 2016), we modeled the accumulated insolation in the region over different periods of time. The three panels in Fig.~\ref{insolation} show the accumulated insolation (in J m$^{-2}$) over one month, half a year, and one year before 25 June 2016, respectively. Results show that the energy received in the Anhur region varies over the area as a result of the local topography. Relatively little sunshine reaches the floor of the canyon structure. However, it is worth noting that the accuracy of the simulated insolation is limited by the accuracy and resolution of the shape model (i.e., ~10 m). 

\section{Discussion}

The Anhur region is highly eroded and exposes some of the deepest layers of the large lobe (Lee et al., 2016; Penasa et al., 2017). 
Evidences of this high erosion rate includes both the high elevation difference between some terraces and the outcropping consolidated terrains (the Anhur and Bes regions together represent the main cliff in the large lobe described by El-Maarry et al (2016)), and the pervasive fracturing of some consolidated regions (e.g., visible in Fig.~\ref{RGBjune2016}), presumably produced by the diurnal and seasonal thermal stress of the material.

The fact that the region exposes inner and more pristine layers is supported by the observations of volatiles. In past studies and this work, several exposures of water ice have been reported since the first observations of the region, starting from March 2015, as well as the first and unique identification of solid CO$_2$ located at the surface. Not only is the region highly eroded, it also shows exposures of different volatiles as well as heterogeneities in the subsurface composition on a scale of tenths of meters. 

Water frost close to the morning shadows was previously observed in the Hapi region during the inbound orbit (de Sanctis et al, 2015) and close to perihelion passage in several regions on both lobes of the comet (Fornasier et al., 2016), with extremely short lifetimes on the order of a few minutes. Water frost was thus related to the diurnal cycle of water (de Sanctis et al, 2015; Fornasier et al., 2016). 

Globally, when pre- and post-perihelion images of regions observed at similar spatial  resolutions and far from perihelion are compared, there is a clear asymmetry in the presence of frost on the comet. Frost is observed mainly post-perihelion, with the notable exception of the Hapi region, where frost was reported at 3 au inbound. Lucchetti et al. (2017) analyzed the evolution of a part of the Seth region in the large lobe and reported some frost or water-ice-enriched regions close to shadows in images acquired in July 2016 (see their Fig. 10), when the comet was at 3.4 au outbound, while there is no evidence of water ice or frost exposure in the same region from the analysis of November 2014 images, when the comet was 3 au inbound. \\
Similarly, the Khonsu region also shows exposure of water ice that locally persists for several months (Deshapryia et al., 2016) during perihelion and post-perihelion observations (Hasselmann et al., 2019, Fornasier et al., 2018, this issue). Some  are related to source regions of jets and/or to important morphological changes (Hasselmann et al., 2019). 

As mentioned before, volatile exposures in the Anhur region were reported since the first observations (Filacchione et al., 2016a; Fornasier et al., 2016, 2017) when the comet was $\sim$ 2 au inbound. However, frost persisting inside the canyon-like structure was first observed in April 2016 and is clearly evident in the high-resolution images acquired during June-July 2016 (Figs.~\ref{RGBjune2016} and ~\ref{30july05}). \\
The higher abundance of frost in the post-perihelion orbit compared to the pre-perihelion orbits may be related to thermal time lag for the turnoff of subsurface volatile sublimation created by the propagation of the thermal wave that is driven by the comet's perihelion passage. 

We must also note that the cometary nucleus changed its color behavior throughout the orbit, becoming progressively bluer while approaching perihelion compared to the inbound
observations at 3 au, followed again by a color reddening at post-perihelion distances $>$ 2 au (Fornasier et al., 2016). This behavior was associated with  the
progressive thinning of the dust mantle when the comet approached
perihelion, exposing the underlying layers enriched in water ice. After perihelion passage, with the progressive decrease of the cometary activity, part of the dust in the coma fell back again onto the nucleus, reddening the surface again. The reddening-bluing-reddening sequence  during pre- in- and post-perihelion observations, respectively, was observed globally on the nucleus (Fornasier et al., 2016) and was also reported for Seth (Lucchetti et al., 2017) and Anhur (Fornasier et al., 2017). Thus, the Anhur region was also partially covered by infalling dust from the coma. Therefore, in addition to the thermal lag effect, the higher frost abundance after perihelion may also indicate that the dust that fell back onto the comet is preserving some water ice, as suggested in Keller et al. (2017). \\
We conclude that frost is preserved in shadows until the internal thermal wave is sufficient to permit volatile recondensation. Some frost may persist at the surface when the comet is far from the Sun (beyond $\sim$ 4 au), when the temperature is low enough (below 80 K, Gulkis et al. 2015) to avoid any water sublimation. However, the nucleus experienced great differences in insolation between the north and south hemisphere because of the orientation of its rotation axis. Several pieces of evidence indicate that the northern hemisphere is covered by back-fall particles that were ejected from the south during the southern summer (Keller et al. 2015, 2017). The northern hemisphere regions are thus progressively covered by a desiccated layer of infalling dust. In the very first resolved observations of Rosetta, when the comet was beyond 3.5 au inbound and when only the northern hemisphere was visible from the spacecraft, frost was rarely observed, with the notable exception of Hapi. It could have been masked by the thick dust deposit, or simply have already sublimated 
considering that water-ice sublimation was the main driver for the cometary activity even at heliocentric distance $>$ 3.5 au (Gulkis et al. 2015).

According to the radio science instrument (RSI) measurements, the estimated total mass loss comparing the measurements at the end and at the beginning of Rosetta observations was about 10$^{10}$ kg (Patzold et al., 2018). However, the amount of uplifted material must have been even greater around perihelion, considering that part of the cometary dust fell back onto the nucleus later, when the activity progressively decreased after the perihelion passage. \\
In Anhur, the main morphological changes for which we were able to provide a mass estimation produce a lower limit in the mass loss of about 5$\times 10^{7}$ kg (including the mass loss associated with the new scarp nearby the Anhur-Bes frontier), that is, $>$0.5\% of the mass loss by the comet. It should be noted, however, that we cannot estimate the mass lost for several observed changes because we are limited by the fact that Anhur, as most of the southern hemisphere regions, was not observed at high spatial resolution before perihelion passage.  The estimate of the mass loss observed in the Khonsu region, where the illumination and observing conditions were more favorable pre-perihelion compared to the available data for Anhur, is between 1.5\% and 4.2\% of the total mass loss o 67P (Hasselmann et al., 2019).

Processes causing the surface changes reported in the literature as well as in this paper are cometary-specific weathering, erosion, and transient events driven by thermal stress and solar insolation, and eventually by other exothermic processes such as water-ice crystallization or clathrate destabilization (El-Maarry et al., 2017; Groussin et al., 2015). Similar processes likely act on other cometary nuclei. Morphological changes were also observed for the 9P/Tempel 1 nucleus between the Deep Impact and Stardust observations. Some smooth areas were noted to have receded by several meters, accompanied by erosion processes of the edges of smooth flows (Veverka et al. 2013; Thomas et al., 2013). These changes have been interpreted to be due to the progressive sublimation and depletion of volatiles and ice-rich material (Meech, 2017), locally detected on the 9P comet surface since the first Deep Impact observations (Sunshine et al., 2006). Tempel 1 experienced a mass loss of 2$\times10^{8}$ kg, which represented 2\% of the one-orbit mass loss for this comet. This loss occurred over 0.01\% of the surface of Tempel 1 (Veverka et al. 2013; Thomas et al., 2013). \\  
In the Anhur region, some of the changes, especially in and close to the canyon structure, were probably driven by the intense activity during the perihelion and post-perihelion passage. The region was  highly active: it was the source of  26 distinct activity events (Vincent et al., 2016a; Fornasier et al., 2017, 2018, this issue). The most intense event is the so-called perihelion outburst originating from the canyon structure, which also hosted other fainter jets that were observed during  June and August 2015; 16 other faint activity events took place during or shortly after perihelion passage (August-beginning of September 2015), some lasted for about one minute (Fornasier et al., 2018); 
5 other activity events were observed in 2016, mostly in 27 January images, including broad-shaped outbursts and an optically thick plume that produced a shadow on the surface, from which Fornasier et al. (2017) estimated an optical depth of $\sim$ 0.43.
 Fornasier et al. (2018, this issue) reported that the activity events they investigated during the cometary southern summer were triggered by illumination conditions and were not associated with a particular terrain type or morphology. The Anhur region behaves in this way as well because activity events are found in different types of terrain, in particular on consolidated terrains and smooth deposits. 

As mentioned before, the source of the perihelion outburst is located within the canyon-like structure (see Figs. 11 and 12 in Fornasier et al., 2018). This huge event, consisting of a collimated jet and a broad structure, may have lifted up the dust coating inside the canyon (Fig.~\ref{areaB}, the surface inside the polygon called A3, and Fig.~\ref{dustbank}) and probably triggered some boulder displacement (Fig.~\ref{areaC}, surface called A2, and boulders B1-B5) and/or the surface changes at the mouth of the canyon (Fig.~\ref{areaC}, feature Z2). Interestingly, the upper limit in the ejected mass of this event (1.8 million kg) is comparable to that of the dust layer removed within the canyon structure.
We estimate for this event a dust production rate of about 300-2000 kg/s, consistent with the rate of emission of large grains in the coma of 67P (Fulle et al., 2016b, 2019). In the coma of 67P, the ejected mass was found to have different particle sizes, but was concentrated in chunks of 10 cm and larger, up to $\sim$ 0.8 m for a boulder observed close to perihelion passage (Fulle et al., 2016b).  

To understand the capacity of an outburst to lift particles to boulders, we need to estimate the water production flux that is released during the event (Hasselmann et al., 2018; El-Maary et al., 2017). Through an OSIRIS image, we can estimate this only indirectly. The water production flux needed to move or lift 50 m sized boulders such as those reported in the Khonsu region (El-Maarry et al., 2017; Hasselmann et al., 2019) is comparable to that of some outbursts observed for comet 67P (Agarwall et al., 2017; Hasselmann et al., 2019). \\
For Anhur, the high insolation during the perihelion passage mainly produced erosion. We have no information about the boulder distribution that could have been removed and/or displaced during the peak of cometary activity because of the relatively poor spatial resolution at and before the perihelion passage. Based on the remaining talus from cliff collapses, we can deduce that some blocks of up to $\sim$ 15 meters (Vincent et al., 2016b; Pajola et al, 2017) could have been displaced during such events, but this remains speculative. \\

In addition to the perihelion outburst event, some faint jets were observed close to the new scarp in region 1 (Fig.~\ref{areaB}) shortly after the perihelion passage, well before its formation in January 2016, and evidence of water-ice exposure observed inside this structure in June-July 2016, pointing to a volatile-enriched subsurface layer. \\
The other sources of faint activity events reported in Fornasier et al. (2018) are sometime close to but not exactly aligned with the morphological changes reported here. It is likely that a number of activity events took place in Anhur and other regions of the cometary resurfacing localized areas such as those observed here, but they were not seen in the Rosetta observations. Most of the events had lifetimes shorter than one hour for the outbursts and as short as one to two minutes for fainter events (Vincent et al., 2016a; Fornasier et al., 2018).

\section{Conclusions}

We analyzed and compared pre- and post-perihelion images of Anhur, finding a number of morphological changes, evidence of several tiny bright patches, and of frost in shadowed regions.
The observed destruction and fragmentation of boulders and development of new cliffs or scarps
suggests that here the cometary surface is weak and very friable. We observed clear evidence of removal of dust mantle in some localized areas of Anhur, especially within the canyon structure, which experienced several changes that were probably driven by the intense perihelion outburst. \\
 The observations of new small-scale exposures of water ice in Anhur near shadowed regions or at the base of new scarps indicate that water ice is very close to the cometary surface and is exposed after recent sublimation. This is typical not only of Anhur, but is also observed elsewhere on the nucleus. Near-surface water ice was exposed after the Aswan (northern hemisphere) cliff collapse (Pajola et al., 2017) and was also observed after a resurfacing of an area with a radius of 10 m in Imhotep (near the cometary equator) that was produced by an outburst in July 2016 (Agarwal et al., 2017). Moreover, the color variations driven by the diurnal and seasonal water cycle (Fornasier et al., 2016) point to ubiquitous water ice in the cometary just below the dust layer. \\
Shadowed areas such as the canyon-like structure act as cold traps and host frost when the comet reaches colder regions of the solar system (beyond 3 au). Frost is ubiquitously observed within the canyon structure in June-July 2016 images and appears as a thin coating resulting from subsurface recondensation of water ice. It is clear from this study that high spatial resolution imagery (submeter) is necessary to detect morphological changes and to deeply study the evolution of cometary surfaces.

\vspace{0.3truecm}
\begin{acknowledgements}

OSIRIS was built by a consortium led by the Max-Planck-Institut f\"ur Sonnensystemforschung, Goettingen, Germany, in collaboration with CISAS, University of Padova, Italy, the Laboratoire d'Astrophysique de Marseille, France, the Instituto de Astrof\'isica de Andalucia, CSIC, Granada, Spain, the Scientific Support Office of the European Space Agency, Noordwijk, The Netherlands, the Instituto Nacional de T\'ecnica Aeroespacial, Madrid, Spain, the Universidad Polit\'echnica de Madrid, Spain, the Department of Physics and Astronomy of Uppsala University, Sweden, and the Institut  f\"ur Datentechnik und Kommunikationsnetze der Technischen Universitat  Braunschweig, Germany. \\ 
The support of the national funding agencies of Germany (DLR), France (CNES), Italy (ASI), Spain (MEC), Sweden (SNSB), and the ESA Technical Directorate is gratefully acknowledged.  We thank the Rosetta Science Ground Segment at ESAC, the Rosetta Mission Operations Centre at ESOC and the Rosetta Project at ESTEC for their outstanding work enabling the science return of the Rosetta Mission. SF acknowledges  the financial  support from the France Agence Nationale de la Recherche (programme Classy, ANR-17-CE31-0004). The authors thank Dr. E. Howell for her comments and suggestions which helped us to improve this article.

\end{acknowledgements}

{}

\begin{appendix}

\section{Supplementary material: Movies}

We provide four movies showing the morphological changes in the four areas of Anhur that are investigated in Figs. 2, 7, 8 and 10:\\
Area1.gif \\
Area2.gif \\
Area3.gif \\
Area4.gif \\

\begin{center}
\begin{figure}
\begin{centering}
\label{perihelionjet}
\includegraphics[scale=0.5]{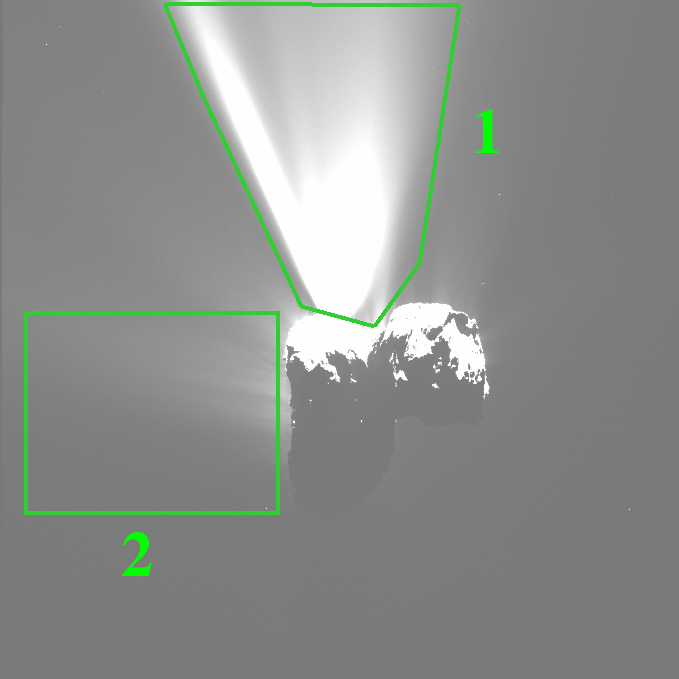}
\par\end{centering}
\caption{Image of the perihelion outburst acquired with the orange filter on 12  August 2015 UT17:20:02. Contrast has been stretched to highlight the
brightness profile of the outburst. The green boxes represent the full bright area
of the event (number 1) and the control area (number 2) that
contains the flux contribution of the quiescent coma between the body
and the spacecraft.}

\end{figure}
\par\end{center}

\end{appendix}


\end{document}